\documentclass[12pt]{article}

\usepackage{graphicx}
\usepackage{amsmath}
\usepackage{amssymb}
\usepackage{wrapfig}
\usepackage{indentfirst}
\usepackage{color}
\usepackage{subfigure}

\begin{document}

\vskip 0.5cm \centerline{\bf\Large Vector meson production}
\centerline{\bf\Large in ultra-peripheral collisions at the LHC}  \vskip 0.3cm
\centerline{R.~Fiore $^{a\clubsuit}$, L.~Jenkovszky $^{b\star}$, V.~Libov$^{c\diamond}$, and Magno V. T. Machado$^{d\spadesuit}$}

\vskip 1cm

\centerline{$^a$ \sl Dipartimento di Fisica, Universit\`a  della Calabria and}
\centerline{Istituto Nazionale di Fisica Nucleare, Gruppo collegato di Cosenza}
\centerline{I-87036 Arcavacata di Rende, Cosenza, Italy}
\centerline{$^b$ \sl Bogolyubov Institute for Theoretical Physics,
National Academy of Sciences of Ukraine} \centerline{\sl Kiev,
03680 Ukraine}
\centerline{$^c$ \sl Deutsches Elektronen-Synchrotron, Hamburg, Germany}
\centerline{$^d$ \sl HEP Phenomenology Group, CEP 91501-970, Porto Alegre, RS, Brazil}
\vskip
0.1cm

\begin{abstract}\noindent
By using a Regge-pole model for vector meson production (VMP), successfully describing the HERA data, we analyse the correlation between VMP cross sections in photon-induced reactions at HERA and those in ultra-peripheral collisions at the Large Hadron Collider (LHC).
Predictions for future experiments on production of $J/\psi$ and $\psi(2S)$ are presented.
\end{abstract}

\vskip 0.1cm

$
\begin{array}{ll}
^{\clubsuit}\mbox{{\it e-mail address:}} &
\mbox{fiore@fis.unical.it}\\
^{\star}\mbox{{\it e-mail address:}} &
   \mbox{jenk@bitp.kiev.ua} \\
^{\diamond}\mbox{{\it e-mail address:}} &
   \mbox{vladyslav.libov@desy.de} \\
 ^{\spadesuit}\mbox{{\it e-mail address:}} &
\mbox{magnus@if.ufrgs.br}\\  

\end{array}
$


\section{Introduction}\label{Int}

Following the shut-down of the HERA collider at DESY (Hamburg), exclusive diffractive production of vector mesons in ultra-peripheral collisions of protons and nuclei became among the priorities of the present and future studies at the CERN LHC \cite{LHCb1, LHCb2, LHC_ATLAS}, triggering a large number of theoretical investigations \cite{Schafer, Brazil, Ryskin, Motyka, Szczurek}.
For a relevant review paper see, e.g., Ref.~\cite{Review}.
First results on vector meson production (VMP), in particular on $J/\psi$, are already published \cite{LHCb1, LHCb2}.

In this study of VMP at the CERN LHC we scrutinize possible changes in the energy dependence of the cross sections when moving from HERA to the LHC, in particular we are interested in the change from light vector mesons to heavy vector mesons ($\phi,\  J/\psi,\  \Upsilon$,  {\it etc.}).

\section{VMP at HERA}
\begin{figure}[!h]
\centering
\includegraphics[width=.8\textwidth]{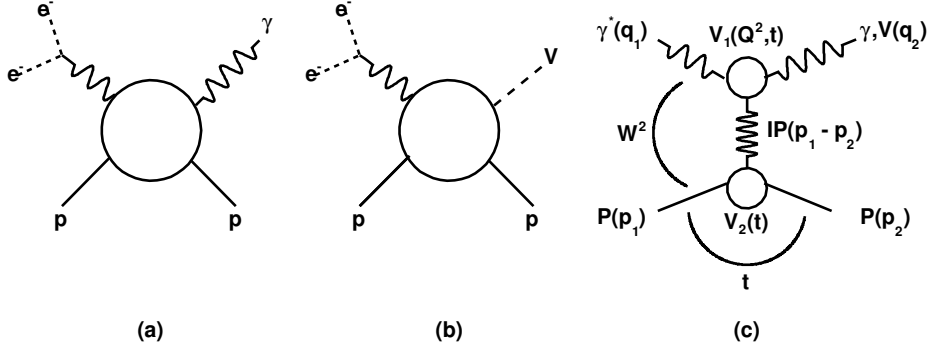}
\caption{Diagrams of DVCS (a) and VMP (b); (c) DVCS (VMP) amplitude in a Regge-factorized form.}
\label{fig:diagrams}
\end{figure}

At HERA, VMP was studied in details both by the H1 and ZEUS collaborations. Most of the events were chosen in the 
kinematical region corresponding to diffractive scattering, which means that the processes can be described by a Pomeron exchange (see Fig. \ref{fig:diagrams}). Pomeron dominance is especially clean in $J/\Psi$ production, where,
by the Zweig (OZI) rule, any exchange of secondary trajectories, made of quarks, is forbidden, thus leaving alone the uncontaminated Pomeron exchange.  This does not mean that the dynamics is simple, but we have the opportunity 
to study in this class of reactions the nature of the Pomeron, a complicated and controversial object. 
The main problem is the twin nature of the Pomeron: it seems to be ``soft" or ``hard" depending on the virtuality of the incident photon and/or the mass of the produced vector meson.   
     
In most of the papers on the subject the existence of two Pomerons is assumed: a hard (or QCD) Pomeron, resulting from 
perturbative quantum chromodynamic calculations, and a soft one, somewhat misleadingly called ``non-perturbative". 
We instead believe that there is only one Pomeron in Nature, but it has two components, whose relative weight is 
regulated by relevant $\widetilde Q^2$-dependent factors in front of the two, where the measure of the ``hardness", $\widetilde Q^2=Q^2+M_V^2$, is the sum of the squared photon virtuality $Q^2$ and the squared mass $M^2_V$ of the produced vector meson.  

A specific model realizing this idea was constructed and tested against the experimental data recently (see
Ref.~\cite{FFJS} and earlier references therein). The relevant VMP amplitude reads 
  $$
   A(s,t,Q^2,{M_v}^2)= \widetilde{A_s}e^{-i\frac{\pi}{2}\alpha_s(t)}\left(\frac{s}{s_{0}}\right)^{\alpha_s(t)}
    e^{b_st - n_s\ln{\left(1+\frac{\widetilde{Q^2}}{\widetilde{Q_s^2}}\right)}}
  $$
  \begin{equation}
  +\widetilde{A_h}e^{-i\frac{\pi}{2}\alpha_h(t)}\left(\frac{s}{s_{0}}\right)^{\alpha_h(t)}
    e^{b_ht - (n_h+1)\ln{\left(1+\frac{\widetilde{Q^2}}{\widetilde{Q_h^2}}\right)}
    +\ln{\left(\frac{\widetilde{Q^2}}{\widetilde{Q_h^2}}\right)} },
    \label{eq:Amplitude_FFJS}
    \end{equation}
 where  $\alpha_s(t)$ and $\alpha_h(t)$ are the soft and hard Pomeron trajectories, $s=W^2$, $W$ being the energy of the VMP. Let us 
stress that the Pomeron is unique in all reactions, but its components (and parameters) vary. 
Examples with detailed fits can be found in the papers of Ref.~\cite{FFJS}.

The integrated (called also total) VMP cross section is given by 
\begin{equation}
\sigma_{\gamma p\rightarrow Vp}(\widetilde Q^2, W)=\int_{t_m}^{t_{thr}}\frac{d\sigma(\widetilde Q^2, W, t)}{dt},
\label{total}
\end{equation} 
where the upper limit is $t_{thr} = 0$ GeV$^2$ and the lower limit is the kinematical one, $t_m = -s/2$. The total cross section 
can be simply calculated, without integration for an exponential diffraction cone, according to the formula
\begin{equation}
\sigma_{el}(s)=\frac{1}{B(s)}\frac{d\sigma}{dt}\biggr\rvert_{t=0},
\label{elastic}
\end{equation}
where $B$ is the forward slope. 

Since our primary goal is the comparison between the energy dependence of VMP at HERA and the LHC, we start with very simple ans\"atz for the $\gamma p\rightarrow Vp$ cross section, postponing the use of the advanced model given by  
Eq. (\ref{eq:Amplitude_FFJS}) to a future study.

\section{VMP at the LHC} 
\subsection{Distribution in rapidity}\label{distribution}
\begin{figure}[!h]
\centering
 \includegraphics[width=.4\textwidth]{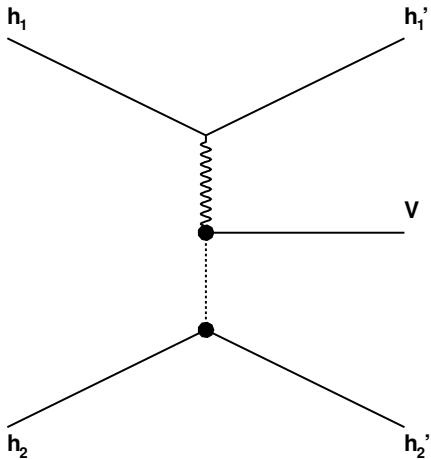}
 \caption{Feynman diagram of vector meson production in a hadronic collision.}
 \label{fig:vmp_feynman}
\end{figure}

Vector meson production (VMP) cross section in hadronic collisions (see the relative Feynman diagram in Fig. \ref{fig:vmp_feynman}) can be written in a factorized form, according to Refs.~ \cite{Brazil, Review} (e.g. Eqs. (1) and (9) in \cite{Brazil}a)).
The distribution in the rapidity $Y$ of the production of a vector meson $V$ in the reaction $h_1+h_2\rightarrow h_1Vh_2,$ (where $h$ may be a hadron, e.g. proton, or a nucleus, pPb, PbPb,...) is calculated according to a standard prescription based on the factorization of the photon flux and photon-proton cross section (see below).

Generally speaking, the $\gamma p$ differential cross section depends on three variables: the total energy   of the $\gamma p$ system, $W$, 
the squared momentum transfer at the proton vertex, $t$, and $\widetilde Q^2=Q^2+M_V^2$, where, as previously said, $Q^2$ is the photon virtuality and $M_V$ is the mass of the produced vector meson. Since, by definition, in ultra-peripheral collisions we have $b>>R_1+R_2$, where $b$ is the impact parameter, i.e. the closest distance between the centres of the colliding particles/nuclei with radii  $R_i$ (i = 1,2),
photons are nearly real, $Q^2 \simeq 0$, and $M_V^2$ remains the only measure of ``hardness". Notice that this might not be true for peripheral collisions, where $b\sim R_1+R_2$, and in the 
Pomeron or Odderon exchange instead of the photon. Finally, the $t$-dependence (shape of the diffraction cone) is known to be nearly exponential. It can be either integrated, or
kept explicit. Extending this parametrization to include a $t-$dependent exponential is easy (see below).
Concerning total cross section $\sigma_{\gamma p\rightarrow Vp}(\widetilde Q^2, t, W)$, it is well known from HERA. 

We use a simple parametrization of the $\sigma_{\gamma p\rightarrow Vp}(W)$ cross section, 
suggested by Donnachie and Landshoff \cite{DL}: $\sigma_{\gamma p\rightarrow Vp}(W)\sim W^{\delta},\ \delta\approx 0.8$ (more involved models, e.g. of Ref. \cite{FFJS, Capua} will be considered elsewhere).

The differential cross section as function of the rapidity $Y$ reads
 $$
\frac{d\sigma (h_1+h_2\rightarrow h_1+V+h_2)}{dY}=\omega_+\frac{dN_{\gamma/h_1}(\omega_+)}{d\omega}\sigma_{\gamma h_2\rightarrow Vh_2}(\omega_+)
$$
\begin{equation}\label{1}
+\omega_-\frac{dN_{\gamma/h_2}(\omega_-)}{d\omega}\sigma_{\gamma h_1\rightarrow Vh_1}(\omega_-),
\end{equation}
where $\frac{dN_{\gamma/h}(\omega)}{d\omega}$ is the ``equivalent" photon flux \cite{Review}, defined as (see, for instance, Ref.~\cite{Brazil} and references therein) 
\begin{equation}
\frac{dN_{\gamma/h}(\omega)}{d\omega} = \frac{\alpha_{em}}{2\pi\omega}[1+(1-\frac{2\omega}{\sqrt{s}})^2]
(\ln\Omega-\frac{11}{6}+\frac{3}{\Omega}-\frac{3}{2\Omega^2}+\frac{1}{3\Omega^3}),
\label{photon_flux}
\end{equation}
and $\sigma_{\gamma p\rightarrow Vp}(\omega)$ is the total cross section of the vector meson photoproduction subprocess (the same as e.g. at HERA, see Ref. \cite{Capua}). 
Here $\omega$ is the photon energy, $\omega=W^2_{\gamma p}/2\sqrt s$, where $\sqrt s$ denotes now the c.m.s. energy of the  proton-proton system; 
$\omega_{min}=M_V^2/(4\gamma_Lm_p),$ where $\gamma_L=\sqrt s/(2m_p)$
is the Lorentz factor (Lorentz boost of a single beam); e.g., for $pp$ at the LHC for $\sqrt{s}=7$~TeV,
$\gamma_L=3731$.
Furthermore,
$\Omega=1+Q_0^2/Q_{min}^2,$ with  $Q_{min}^2=(\omega/\gamma_L)^2$ and  $Q_0^2=0.71 GeV^2$,  
$Y\sim\ln(2\omega/M_V)$.  
The differential cross section in $W$ can be calculated as $d\sigma/dW=(d\sigma/dY)/W$.

For definiteness we fix: a) the colliding particles are protons;
b) the produced vector meson $V$ is $J/\psi$, and 3) the collision energy $\sqrt s=7$ TeV.
We comprise the constants in $A=\alpha_{em}/(2\pi),\ \  c=Q_0^2\gamma_L^2$.
(Note that the shape of the distribution in $Y$ is very sensitive to the value and the sign of the constant $c$).
The $i=\pm$ signs of $\omega$ correspond to the first or second term in Eq. (\ref{1}), respectively, $\omega_{\pm}\sim e^{\pm Y}$.

\subsection{Corrections for rapidity gap survival probabilities}\label{corrections}
The above results may be modified by initial and final state interactions,
alternatively called as rescattering corrections. The calculation of these
corrections is by far not unambiguous, the result depending both on the input
and on the unitarization procedure chosen. The better (more realistic) the input, the smaller the unitarity (rapidity gap survival probability) corrections.
Since this is a complicated and controversial issue {\it per se} deserving special studies beyond the scope of the present paper, to be coherent with the ``common trend", here we use only familiar results from the literature:
the standard prescription is to multiply the scattering amplitude (or the cross section) by a factor (smaller than one), depending on the energy and eventually other kinematical variables~\cite{Ryskin}.
In this work we used the constant value of 0.8.

\section{Results of the fitting procedure}

In this Section theoretical predictions are presented and compared to experimental data.
IN our alculations we use two models for the exclusive $J/\Psi$ photoproduction cross section, $\sigma_{\gamma p \to J/\Psi}$:
the simple power-law, $\sigma_{\gamma p \to J/\Psi}\sim W^\delta$, with $\delta=0.8$, and the so-called Reggeometric model.
The latter, in its first version, was suggested and applied to deeply virtual Compton scattering (DVCS) in Ref.~\cite{Capua}.
Apart from $W$ and $t$, it contains also the dependence on virtuality, $Q^2$.
The model was fitted to the HERA data on DVCS, but it can be applied also to the VMP by refitting its parameters.
In this paper we use the first (simpler) version of the Reggeometric model of VMP and DVCS, suggested in Ref. \cite{FFJS}.

\subsection{Fitting the  Reggeometric model to LHCb data}

The first version of the Reggeometric model \cite{FFJS} a) applies to photoproduction ($Q^2=0$). The total photoproduction cross section, Eq. (11) in Ref. \cite{FFJS} a), is given by 
\begin{equation}
\sigma_{\gamma p \to J/\Psi p}=A_0^2\frac{(W/W_0)^{4(\alpha_0-1)}}{(1+\tilde Q^2/Q^2_0)^{2n}[4\alpha'\ln(W/W_0)+4\Bigl(\frac{a}{\tilde Q^2}+\frac{b}{2m_N^2}\Bigr)]},
\end{equation}
where $\widetilde Q^2=Q^2+M_V^2$. The parameters, fitted \cite{FFJS} a) to the $J/\psi$ photoproduction, quoted in
Table II of Ref. \cite{FFJS} a), are: $A_0=29.8\pm 2.8,\ \ Q_0^2=2.1\pm 0.4,\ \
n=1.37\pm 0.14,\ \ \alpha_0 =1.20\pm 0.02,\ \ \alpha'=0.17\pm 0.05, a=1.01\pm 0.11,\ \ b=0.44\pm 0.08,\ \ W_0=1$ and the relevant dimensions here again are implied.
Note that compared to the original formula, $s$ was replaced by $W^2$, since $W$ is used in this paper to 
denote the photon-proton centre-of-mass energy (in contrast, $\sqrt{s}$ in this paper is the proton-proton centre-of-mass energy).

\begin{figure}[!t]
  \centering
  \subfigure[]{\label{fig:dSigma_dy_nodata}
    \includegraphics[height=6.5cm,width=0.46\textwidth]{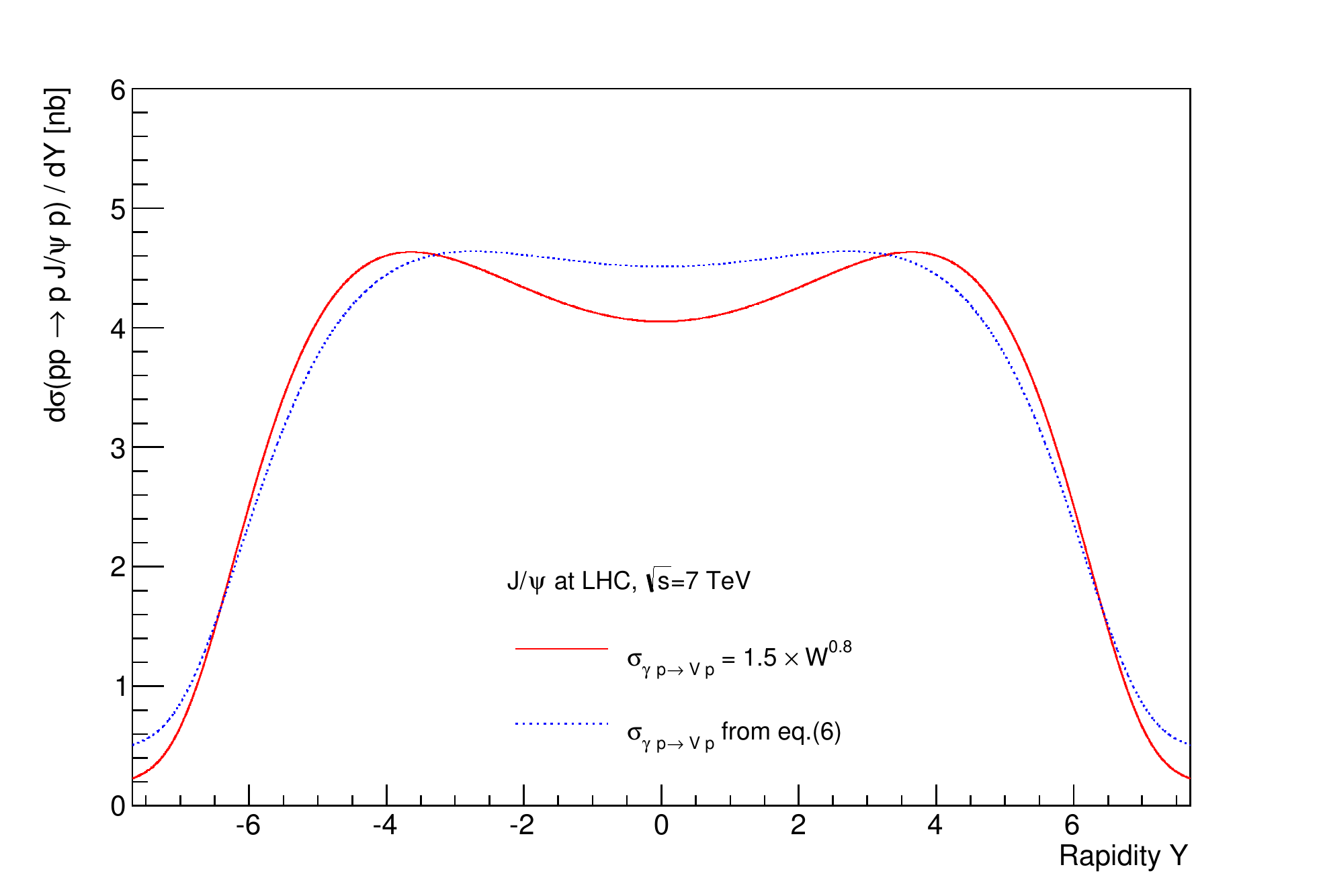}
  }
  \subfigure[]{\label{fig:dSigma_dW_nodata}
    \includegraphics[height=6.5cm,width=0.46\textwidth]{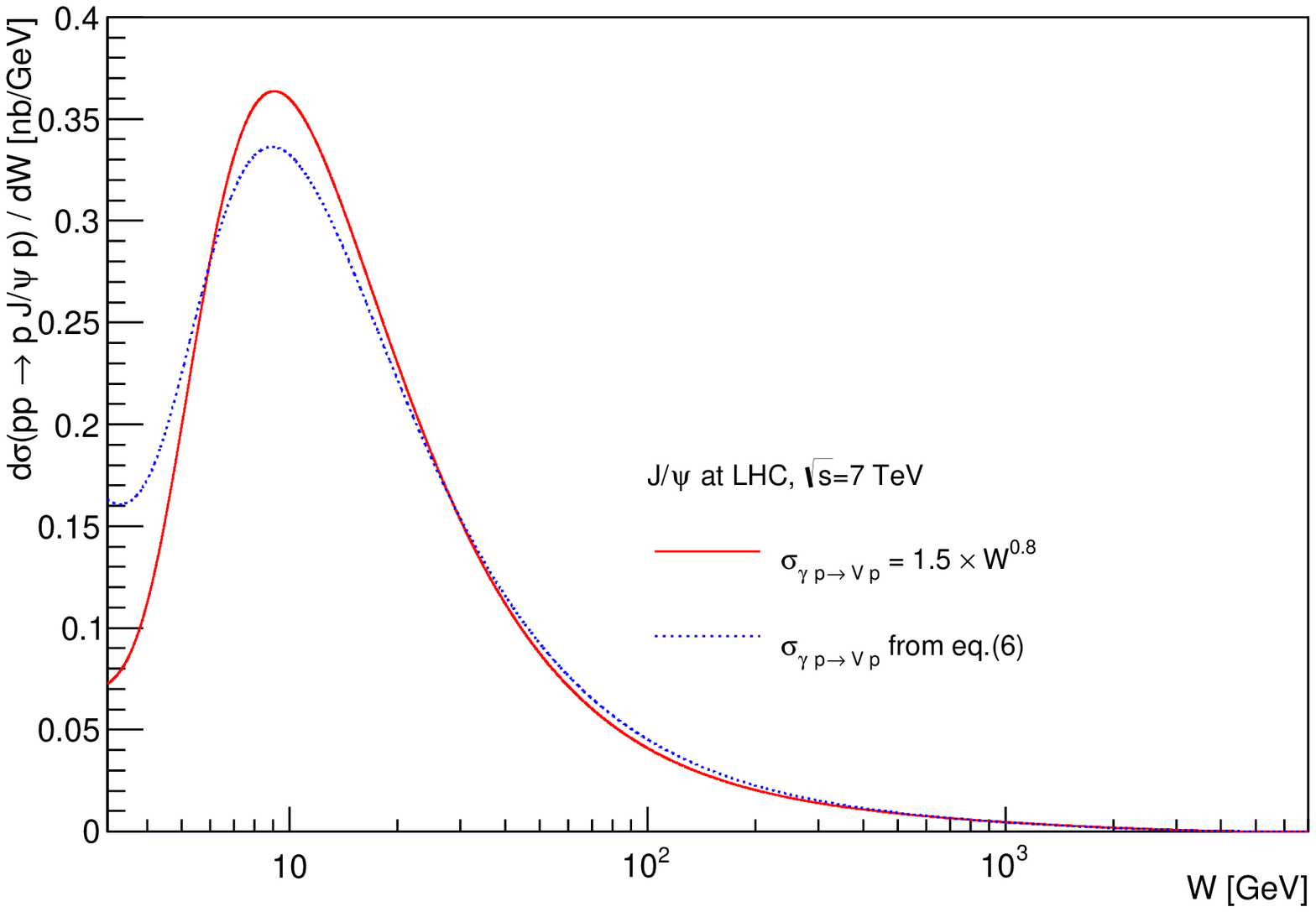}
  }
  \caption{Differential cross section of exclusive $J/\Psi$ production at the LHC as function of the rapidity $Y$, plot (a), and of the photon-proton c.m.s. energy $W$, plot (b).
           The red curves and the blue curves represent, respectively,  the parameterization of the photon-proton cross section according to the simple power-law and  the Reggeometric model.}
\end{figure}

Fig.~\ref{fig:dSigma_dy_nodata} shows the predicted differential cross section of $J/\Psi$ production at LHC ($\sqrt s=7$ TeV) as a distribution in rapidity $Y$.
On the whole, the power law and the Reggeometric model yield similar distributions,  the latter is flatter though.
Fig.~\ref{fig:dSigma_dW_nodata} shows the $W$-dependence of $J/\psi$ photoproduction differential cross section.
Again, the two models give generally similar behaviours.

\begin{figure}[!h]
  \centering
  \subfigure[]{\label{fig:dSigma_dy_comparison1}
    \includegraphics[height=6.5cm,width=0.46\textwidth]{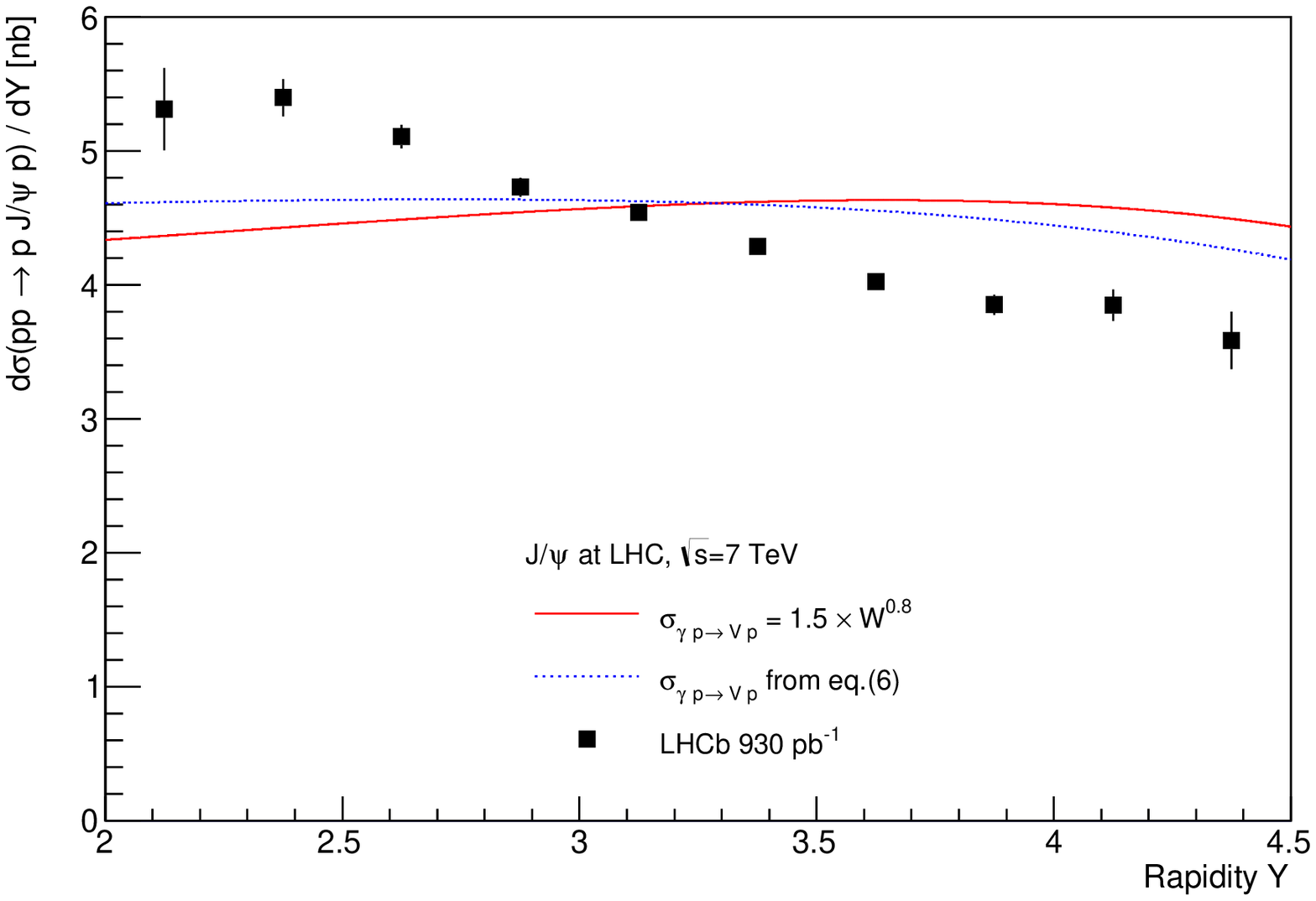}
  }
  \subfigure[]{\label{fig:sigma_gamma_p_W_powerlaw_vs_geometric}
    \includegraphics[height=6.5cm,width=0.46\textwidth]{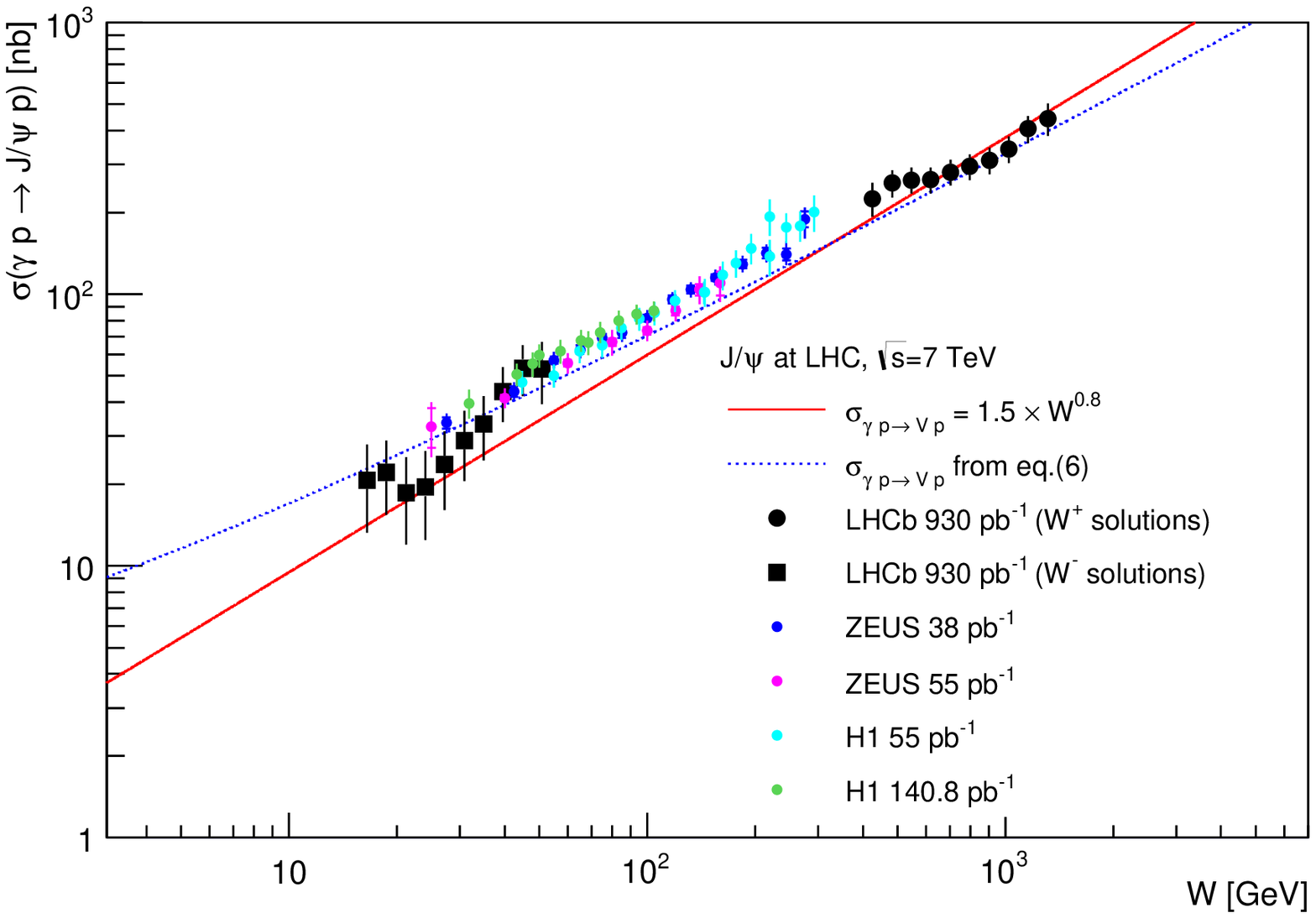}
  }
  \caption{The differential cross section of $J/\Psi$ production at LHC as a function of the rapidity $Y$ together with the LHCb data is shown in plot (a). Also shown, in plot (b), is the $J/\Psi$ photoproduction ($\gamma p \to J/\Psi p$) cross section as a function of the photon-proton c.m.s. energy compared
           with the LHCb, ZEUS and H1 data.}
\end{figure}

The LHCb collaboration at the LHC has recently~\cite{LHCb1, LHCb2} measured the production cross section of $J/\Psi$ as a function of the rapidity Y.
Fig.~\ref{fig:dSigma_dy_comparison1} shows a comparison of our calculations with these data.
As it can be seen, the data are somewhat steeper than both the curves.
LHCb also extracted the basic photon-proton photoproduction cross section as a function of $W$ from the data.
The result is compared to our predictions in  Fig.~\ref{fig:sigma_gamma_p_W_powerlaw_vs_geometric}.
Also the ZEUS and H1 data are shown.

\subsection{Fitting the power law to the LHCb data}

\begin{figure}[!h]
\centering
 \includegraphics[width=.8\textwidth]{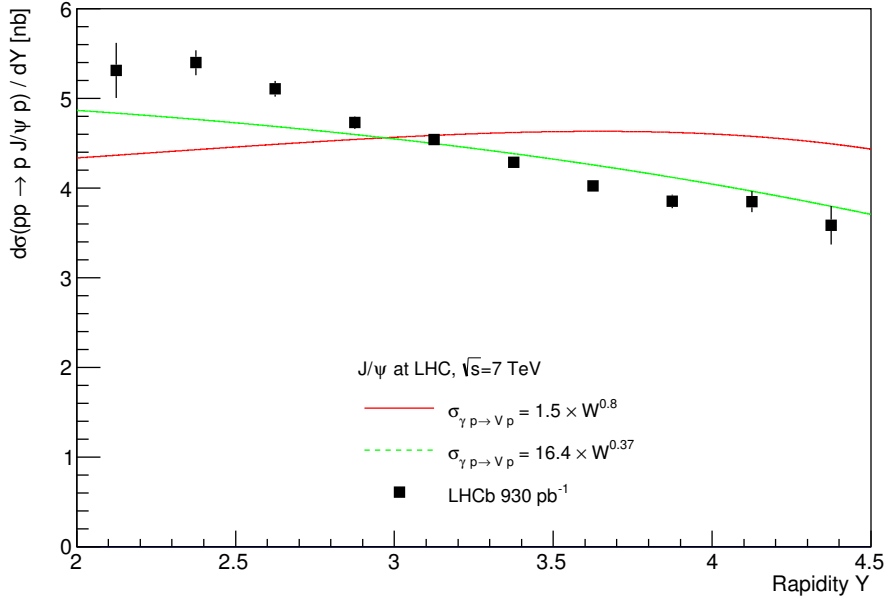}
 \caption{Differential cross section of $J/\psi$ production at the LHC as function of the rapidity $Y$, according to the power law  together with LHCb data.
          Both calculations use the simple power law model, with $\delta=0.8$ (red line) and $\delta=0.37$ (green line).}
 \label{fig:delta_fit}
\end{figure}

\begin{figure}[p]
\centering
 \includegraphics[width=.8\textwidth]{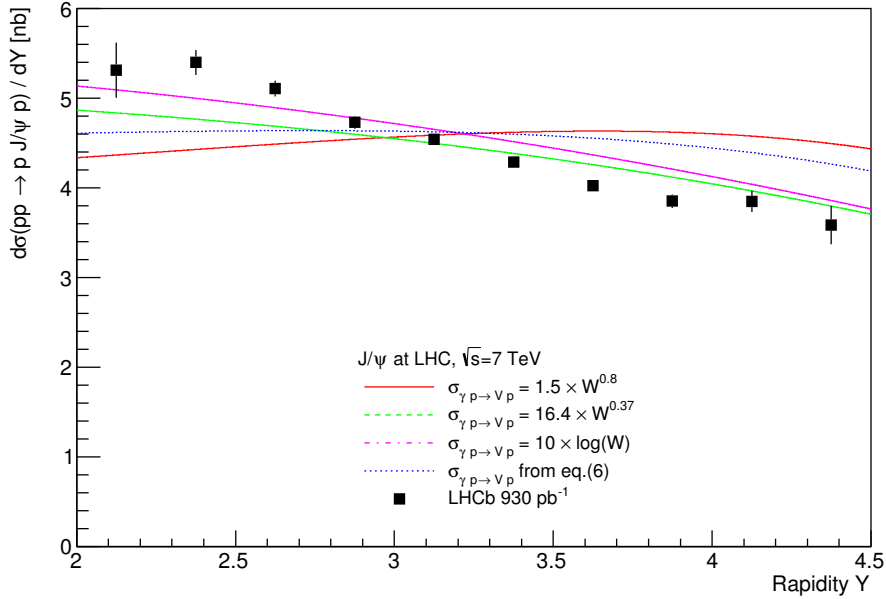}
 \caption{Differential cross section of $J/\psi$ production at the LHC as a function of rapidity $Y$, according to the various models for the photon-proton cross section used: power law with $\delta=0.8$ and $\delta=0.37$ (red and green lines, respectively), logarithmic parameterization (pink line) and the Reggeometric model (blue line). Also the LHCb data are shown.}
  \label{fig:all_curves}
\end{figure}

\begin{figure}[p]
\centering
 \includegraphics[width=.8\textwidth]{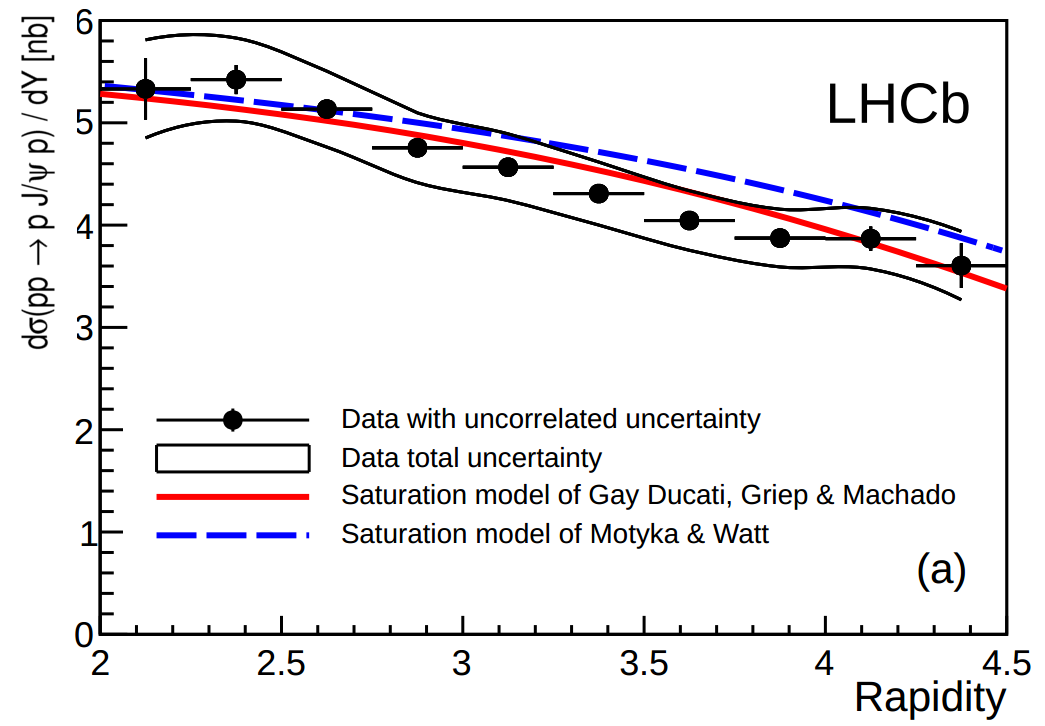}
 \caption{Differential cross section of $J/\psi$ production at LHC as a function of rapidity $Y$ together with the LHCb data \cite{LHCb2}.}
 \label{fig:dipole}
\end{figure}

As discussed above, the LHCb rapidity data have a steeper shape than our predictions.
Here we investigate the role of the power $\delta$ in the rapidity cross section.
In particular, we perform a least-square fit to these data.
The free parameters are the power $\delta$ and the overall normalization.
The best fit gives $\delta$=0.37 and is shown in Fig.~\ref{fig:delta_fit}.
Indeed, the prediction  obtained with this value gives a much better description of the rapidity distribution.
It is interesting make a comparison with other parameterizations, in particular by using the
logarithmic parameterization, which leads to a less steep line than that obtained in the power law case .
This is shown  in Fig.~\ref{fig:all_curves}. The logarithmic parameterization provides a good description as well.
Fig.~\ref{fig:dipole} shows, for completeness, the predictions obtained from other groups, together with the LHCb rapidity data.
The agreement with the data is reasonable.

\begin{figure}[!t]
\centering
 \includegraphics[width=.8\textwidth]{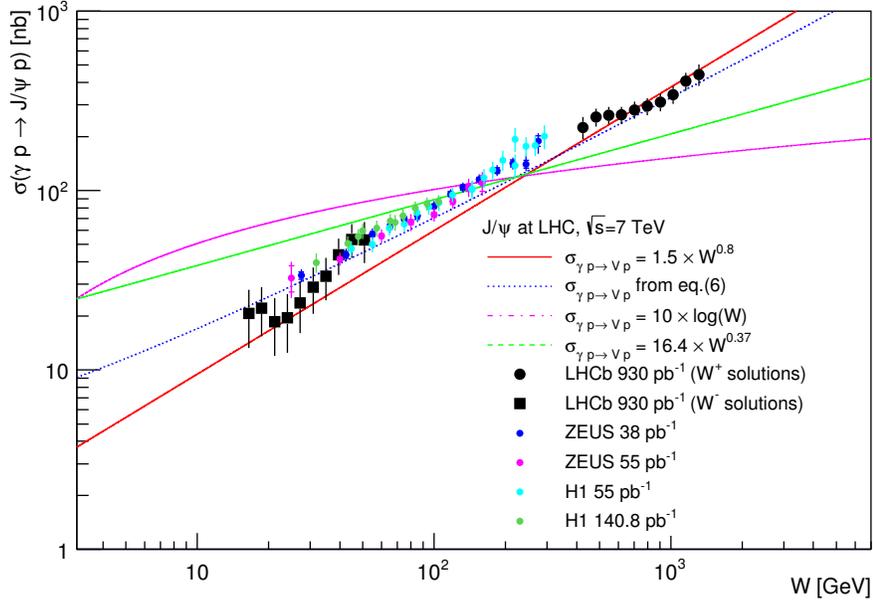}
 \caption{$J/\psi$ photoproduction ($\gamma p \to J/\Psi p$) cross section as a function of the photon-proton c.m.s. energy compared
           with the LHCb, ZEUS and H1 data. Other details are as in Fig.~\ref{fig:all_curves}.}
 \label{fig:all_curves_w}
\end{figure}

Our predictions for the $J/\Psi$ photoproduction, also compared to the H1 and ZEUS data, are shown in Fig.~\ref{fig:all_curves_w}.
As it can be seen, the Reggeometric model provides the best description.
The power law with the standard $\delta=0.8$ agrees reasonably as well.
However, the one with $\delta=0.37$, as well as the logarithmic parameterization fail to describe the data.

\begin{figure}[!ht]
\centering
 \includegraphics[width=.8\textwidth]{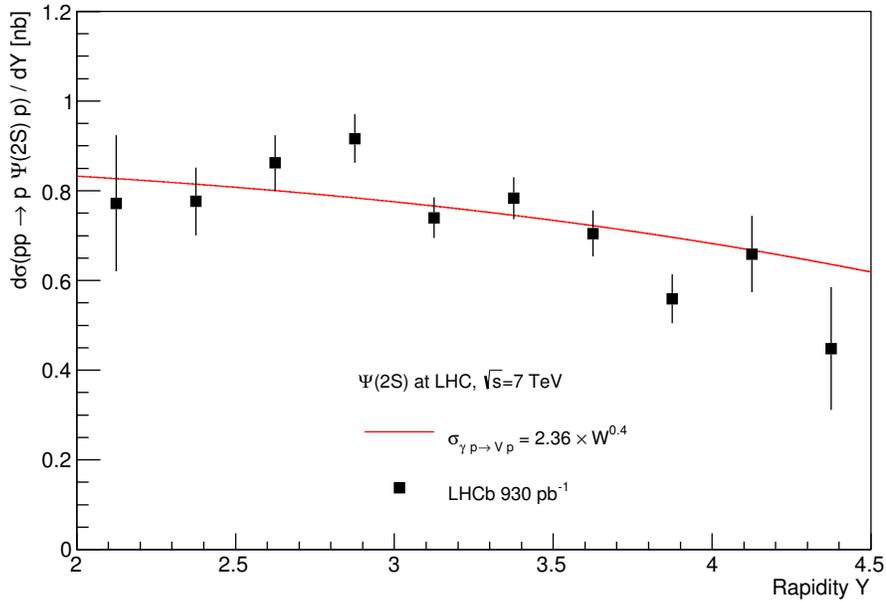}
 \caption{$\Psi(2S)$ differential cross section as a function of rapidity $Y$ compared with the LHCb data. The power law model was used with parameters as shown in the figure.}
 \label{fig:psi2s}
\end{figure}

$\Psi(2S)$ differential cross section as a function of rapidity $Y$ is compared with the LHCb data in Fig.~\ref{fig:psi2s}. The power law model was used with parameters as obtained from the fit. Good description is observed.

\clearpage

\section{Conclusions and Outlooks}

In this paper predictions for exclusive $J/\Psi$ meson production in ultra-peripheral collisions at the LHC are presented and compared with the recent experimental data collected by the LHCb.
The HERA  experimental data are also shown and compared with the H1 and ZEUS measurements.
The simple power-law and a more advanced geometric model, the so called Reggeometric model, were used to describe the photon-proton cross section.
The LHCb for the cross section as a function of the rapidity are steeper than predictions. A better description can be obtained by tuning the power,
however this makes the predictions inconsistent with the HERA data. 

The present study will be extended by
\begin{itemize}
\item the inclusion of the $t$ dependence of the differential cross section - both exponential (corresponding to linear Regge trajectories) and with deviations from the exponential 
(non-linear Regge trajectories);
\item the inclusion of the dependence of the $\sigma_{\gamma p \rightarrow V p}$ cross section on $Q^2$, negligible in $\gamma-$ but important in the case of Reggeon (Pomeron, Odderon,...) exchanges;
\item studies of inelastic processes, i.e. those in which additional particles are produced due to either gluon radiation and/or (c,d) proton dissociation;
\item more advanced studies of corrections due to the rapidity gap survival probability;
\item generalizations to include nuclear collitions.
\end{itemize}


 \section*{Acknowledgements}
We thank A. Salii for discussions. L.J. is grateful to the Dipartimento di Fisica dell'Universit`a della Calabria and the Istituto Nazionale di Fisica Nucleare - Gruppo Collegato di Cosenza, where part of this work was done, for their hospitality and support. 
He was supported also by the grant "Matter under extreme conditions" of the
National Academy of Sciences of Ukraine, Dept. of Astronomy and
Physics and by the DOMUS Curatorium of the Hungarian Academy of Sciences.

\end{document}